\documentclass[smallextended]{svjour3}       
\smartqed  

\RequirePackage[OT1]{fontenc}

\usepackage{amssymb,mathrsfs,lineno,adjustbox,tabularx,ragged2e,slashbox,float, physics}

\usepackage{multirow}
\usepackage{geometry}

\usepackage{amsmath}
\usepackage{algorithm}
\usepackage{algorithmic}
\usepackage{parselines}
\usepackage{epsfig}
\usepackage[nottoc]{tocbibind}
\usepackage[colorlinks,citecolor=blue,urlcolor=blue,filecolor=blue,backref=page]{hyperref}
\usepackage{graphicx}
\usepackage{graphics}
\usepackage{subfigure}
\usepackage{verbatim}
\usepackage{float}
\usepackage{mathtools}
\usepackage{enumerate}
\usepackage{natbib}
\usepackage{url}

\usepackage{mathptmx}      
\begin{document}

\title{Parameter Estimation of four parameter absolute continuous Geometric Marshall-Olkin bivariate Pareto Distribution} 


\titlerunning{EM-BVPA}        
\author{Biplab Paul, Arabin Kumar Dey, Arjun K Gupta and Debasis Kundu\\
}

           

\institute{B. Paul \at
             Department of Mathematics,\\ 
             IIT Guwahati,\\   
             Guwahati, India\\
             \email{biplab.paul@iitg.ac.in}  \\ 
\\
            A. K. Dey \at
              Department of Mathematics, \\
              IIT Guwahati,\\
              Guwahati, India\\
              Assam\\
              Tel.: +91361-258-4620\\
              \email{arabin@iitg.ac.in}\\ 
\\
            Arjun K Gupta \at
             Department of Mathematics and Statistics,\\ 
             Bowling Green State University, USA\\         
             \email{gupta6731@gmail.com} 
             \and
            Debasis Kundu \at
             Department of Mathematics,\\ 
             IIT Guwahati,\\   
             Guwahati, India\\
             \email{kundu@iitk.ac.in}            
}



\maketitle

\begin{abstract}

  In this paper we formulate a four parameter absolute continuous Geometric Marshall-Olkin bivariate Pareto distribution and study its parameter estimation through EM algorithm and also explore the bayesian analysis through slice cum Gibbs sampler approach.  Numerical results are shown to verify the performance of the algorithms.  We illustrate the procedures through a real life data analysis.

\end{abstract}


\section{Introduction}

  Geometric absolute continuous Marshall-Olkin bivariate Pareto (GBBVPA) is more flexible model than absolute continuous Marshall-Olkin bivariate Pareto.  The distribution can be used to model data related to finance, insurance, environmental sciences and internet network.  In this paper we explore the statistical inference of GBBVPA through EM algorithm.  We also study bayesian analysis in this set up through slice cum Gibbs sampler approach.  One of the required criteria for a data set which is to be used for modeling with GBBVPA should not have any equal components. 

  \cite{KozubowskiPanorska:2008} and \cite{Barreto:2012}, introduced geometric bivariate exponential and geometric bivariate gamma distributions, respectively, along the same line.  Series of papers can also be found on statistical inferences on different distributions in the work of Kundu et al. [\cite{Kundu:2017}, \cite{KunduNekoukhou:2018} and \cite{KunduGupta:2014}].  Few recent paper of Asimit et al [\cite{AsimitFurmanVernic:2016}, \cite{AsimitFurmanVernic:2013}, \cite{AsimitFurmanVernic:2010}] and \cite{DeyPaul:2017} discussed statistical inference of singular bivariate Pareto with location scale parameter and its applications.  Recently Dey and Paul [\cite{DeyPaul:2017}] also studied singular four parameter Geometric Marshall Olkin bivariate Pareto distribution.  However there is no paper available in statistical inference on absolute continuous Geometric Marshall-Olkin bivariate Pareto distribution.  We also explore the bayesian analysis under informative prior.  Both frequentist and bayesian confidence intervals are provided along with a illustrative real-life data example.  

 Maximum likelihood estimate may not only be computationally expensive, but also creat problems in finding suitable initial guesses.  To resolve the issues we construct an EM algorithm.  We also explore bayesian approach through slice cum gibbs sampler.  Usual step-out procedure works quite well and easy to implement.  Since it is a very flexible model when components are not equal, it gives the practitioner a choice of an alternative bivariate Pareto model, which may provide a better fit than existing Marshall-Olkin bivariate Pareto distribution. 

  Rest of paper is arranged as follows.  In section 2 we show the formulation of Marshall-Olkin bivariate Pareto distribution. Section 3 discusses Maximum likelihood estimation through EM algorithm.  Bayesian analysis is discussed in Section 4.  Numerical results are shown in Section 5.  Data analysis is shown for all methods in section 6.  We conclude the paper in section 7.

\section{Formulation of Block-Basu bivariate Pareto Geometric Distribution}

\subsection{Brief of singular Geometric bivariate Pareto Distribution}

 A bivariate random variable $(X_1, X_2)$ is said to be distributed according to Marshall-Olkin bivariate Pareto distribution i.e., $(X_1, X_2) \sim MOBVPA(\mu_1, \mu_2, \sigma_1, \sigma_2, \alpha_0, \alpha_1, \alpha_2)$ if it has the cumulative survival function
\begin{align*}
S(x_1, x_2)&= (1+z)^{-\alpha_0}\Big(1 + \frac{x_1-\mu_1}{\sigma_1}\Big)^{-\alpha_1}\Big(1 + \frac{x_2-\mu_2}{\sigma_2}\Big)^{-\alpha_2}\\&=
\begin{cases}
S_1(x_1, x_2), \quad \text{if $\frac{x_1-\mu_1}{\sigma_1}$ \textless $\frac{x_2-\mu_2}{\sigma_2}$}
\\
S_2(x_1, x_2), \quad \text{if $\frac{x_1-\mu_1}{\sigma_1}$ \textgreater $\frac{x_2-\mu_2}{\sigma_2}$}\\
S_{0}(x), \quad \text{if $\frac{x_1-\mu_1}{\sigma_1}$ = $\frac{x_2-\mu_2}{\sigma_2}$}=x
\end{cases}
\end{align*}
where 
\begin{align*}
S_1(x_1, x_2)&=  \Big(1 + \frac{x_2-\mu_2}{\sigma_2}\Big)^{-\alpha_0-\alpha_2} \Big(1 + \frac{x_1-\mu_1}{\sigma_1}\Big)^{-\alpha_1}\\
S_2(x_1, x_2)&=  \Big(1 + \frac{x_2-\mu_2}{\sigma_2}\Big)^{-\alpha_2}\Big(1 + \frac{x_1-\mu_1}{\sigma_1}\Big)^{-\alpha_0-\alpha_1}\\
S_0(x)&= (1 + x)^{-\alpha_0-\alpha_1-\alpha_2}
\end{align*}

so it's joint pdf can be written as 
\begin{align}
\label{2e1}
f(x_1, x_2)&=
\begin{cases}
f_1(x_1, x_2),\quad \text{if $\frac{x_1 - \mu_1}{\sigma_1}$ \textless $\frac{x_2 - \mu_2}{\sigma_2}$}
\\
f_2(x_1, x_2),\quad \text{if $\frac{x_1 - \mu_1}{\sigma_1}$ \textgreater $\frac{x_2 - \mu_2}{\sigma_2}$}\\
f_{0}(x), \quad \text{if $\frac{x_1 - \mu_1}{\sigma_1}$ = $\frac{x_2 - \mu_2}{\sigma_2}$}=x 
\end{cases} 
\end{align}
where 
\begin{align*}
f_1(x_1, x_2)&=\frac{\alpha_1 (\alpha_0 + \alpha_2)}{\sigma_1 \sigma_2}\Big(1 + \frac{x_2-\mu_2}{\sigma_2}\Big)^{-\alpha_0-\alpha_2-1}\Big(1+\frac{x_1-\mu_1}{\sigma_1}\Big)^{-\alpha_1-1}\\
f_2(x_1, x_2)&=\frac{\alpha_2 (\alpha_0 + \alpha_1)}{\sigma_1 \sigma_2}\Big(1 + \frac{x_2-\mu_2}{\sigma_2}\Big)^{-\alpha_2-1}\Big(1+\frac{x_1-\mu_1}{\sigma_1}\Big)^{-\alpha_0-\alpha_1-1} \\
f_0(x)&= \alpha_0(1 + x)^{-\alpha_0-\alpha_1-\alpha_2-1}
\end{align*}

 Suppose $\{(X_{1i}, X_{2i}): i = 1, 2, \cdots\}$ is a sequence of iid bivariate random variables with same cdf $F(\cdot, \cdot)$ and pdf $f(\cdot, \cdot)$.  $N$ is an univariate random variable independent with $(X_{1n}, X_{2n})$'s follow Geometric distribution with $0 < \theta \leq 1$.  We also consider that $(Y_1, Y_2)$ is another bivariate random variable defined as,
\begin{center}
$Y_1 = \text{min}\{X_{11}, X_{12}, \cdots, X_{1N}\}$\\
$Y_2 = \text{min}\{X_{21}, X_{22}, \cdots, X_{2N}\}$
\end{center}
The joint cumulative survival function can be obtained as,
\begin{align}
\label{2e2}
\bar{G}(y_1, y_2)=\frac{\theta S(y_1, y_2)}{1 - (1 - \theta)S(y_1, y_2)} 
\end{align}
So it's a joint density function that can be presented as,
\begin{align}
\label{2e3}
g(y_1, y_2) = \frac{\theta\{1 - (1 - \theta)S(y_1, y_2)\}\pdv[n]{S(y_1, y_2)}{y_1}{y_2} + 2\theta(1-\theta)\pdv{S(y_1, y_2)}{y_1}\pdv{S(y_1, y_2)}{y_2}}{\{1 - (1 - \theta)S(y_1, y_2)\}^3}
\end{align}

  Now we use the survival function of MOBVPA in equation-\ref{2e3}, then the joint distribution of $(Y_1, Y_2)$ is called Geometric Marshall-Olkin bivariate Pareto (G-MOBVPA) distribution.  Therefore the joint survival function of $(Y_1, Y_2)$ can be written as,
\begin{align}
\label{2e4}
\bar{G}(y_1, y_2) &=
\begin{cases}
\bar{G}_1(y_1, y_2), \quad \text{if $\frac{y_1 - \mu_1}{\sigma_1}$ \textless $\frac{y_2 - \mu_2}{\sigma_2}$}
\\
\bar{G}_2(y_1, y_2), \quad \text{if $\frac{y_1 - \mu_1}{\sigma_1}$ \textgreater $\frac{y_2 - \mu_2}{\sigma_2}$}\\
\bar{G}_0(y), \quad \text{if $\frac{y_1 - \mu_1}{\sigma_1}$ = $\frac{y_2 - \mu_2}{\sigma_2}$}\\
\end{cases}
\end{align}
where 
\begin{align*}
\bar{G}_1(y_1, y_2)&= \frac{\theta\big(1 + \frac{y_1 - \mu_1}{\sigma_1}\big)^{-\alpha_1}\big(1 + \frac{y_2-\mu_2}{\sigma_2}\big)^{-\alpha_0-\alpha_2}}{1 - (1 - \theta)\big(1 + \frac{y_1 - \mu_1}{\sigma_1}\big)^{-\alpha_1}\big(1 + \frac{y_2-\mu_2}{\sigma_2}\big)^{-\alpha_0-\alpha_2}} \\
\bar{G}_2(y_1, y_2)&= \frac{\theta\big(1 + \frac{y_1 - \mu_1}{\sigma_1}\big)^{-\alpha_0-\alpha_1}\big(1 + \frac{y_2 - \mu_2}{\sigma_2}\big)^{-\alpha_2}}{1 - (1 - \theta)\theta\big(1 + \frac{y_1-\mu_1}{\sigma_1}\big)^{-\alpha_0-\alpha_1}\big(1 + \frac{y_2 - \mu_2}{\sigma_2}\big)^{-\alpha_2}} \\
\bar{G}_0(y)&= \frac{\theta (1 + y)^{-\alpha_0 - \alpha_1 - \alpha_2}}{1 - (1 - \theta)(1 + y)^{-\alpha_0 - \alpha_1 - \alpha_2}}\\
\end{align*}

Hence the joint pdf is,
\begin{align}
\label{2e.1}
g(y_1, y_2)&=
\begin{cases}
g_1(y_1, y_2),\quad \text{if $\frac{y_1 - \mu_1}{\sigma_1}$ \textless $\frac{y_2 - \mu_2}{\sigma_2}$}
\\
g_2(y_1, y_2),\quad \text{if $\frac{y_1 - \mu_1}{\sigma_1}$ \textgreater $\frac{y_2 - \mu_2}{\sigma_2}$}\\
g_{0}(y),  \quad \text{if $\frac{y_1 - \mu_1}{\sigma_1}$ = $\frac{y_2 - \mu_2}{\sigma_2}$}=y 
\end{cases} 
\end{align}
where 
\begin{align*}
g_1(y_1, y_2)&=\frac{\theta\alpha_1(\alpha_0 + \alpha_2)\big(1 + \frac{y_1 - \mu_1}{\sigma_1}\big)^{-\alpha_1-1}\big(1 + \frac{y_2 - \mu_2}{\sigma_2}\big)^{-\alpha_0 - \alpha_2 - 1}\{1 + (1 - \theta)S_1(y_1, y_2)\}}{\sigma_1 \sigma_2\{1 - (1 - \theta)S_1(y_1, y_2)\}^3}\\
g_2(y_1, y_2)&=\frac{\theta(\alpha_0 + \alpha_1)\alpha_2\big(1 + \frac{y_1 - \mu_1}{\sigma_1}\big)^{-\alpha_0 - \alpha_1 - 1}\big(1 + \frac{y_2 - \mu_2}{\sigma_2}\big)^{- \alpha_2 - 1}\{1 + (1 - \theta)S_2(y_1, y_2)\}}{\sigma_1\sigma_2\{1 - (1 - \theta)S_2(y_1, y_2)\}^3} \\
g_0(y)&= \frac{\theta\alpha_0(1 + y)^{-\alpha_0 - \alpha_1 - \alpha_2 - 1}}{\{1 - (1 - \theta)(1 + y)^{-\alpha_0 - \alpha_1 - \alpha_2}\}^2}
\end{align*}

  We denote this distribution as $G-MOBVPA(\theta, \mu_1, \mu_2, \sigma_1, \sigma_2, \alpha_0, \alpha_1, \alpha_2)$.  From Lebesgue decomposition theorem, the joint pdf of $(Y_1, Y_2)$ can be written as,
 
\begin{align}
\label{2e.2}
g(y_1, y_2)&=\frac{\alpha_1 + \alpha_2}{\alpha_{0} + \alpha_1 + \alpha_2}g_{ac}(y_1, y_2) + \frac{\alpha_0}{\alpha_0 + \alpha_1 + \alpha_2}g_{s}(y)
\end{align} 
where $g_{ac}(y_1, y_2)$ and $g_s(y_1, y_2)$ are the absolute continuous part and singular part of G-MOBVPA distribution.  Here we are interested in absolute continuous part only.

\subsection{Absolute continuous Geometric Marshall-Olkin bivariate Pareto distribution (G-BBBVPA)}

\begin{figure}[H]
	\centering
	\includegraphics[height=2in, width=2.5in]{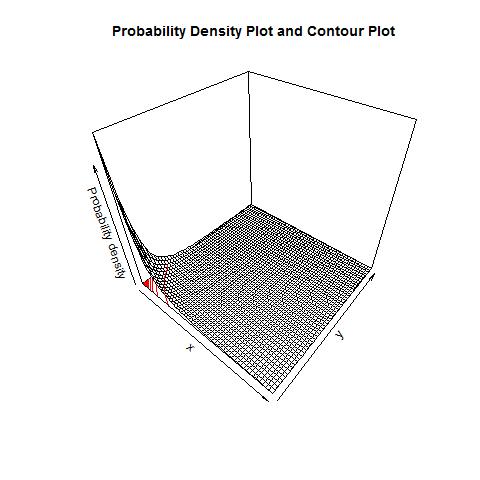}
	\caption{Probability density plot and Contour Plot for the parameter values $\theta = 0.20$, $\alpha_0=0.10$, $\alpha_1=0.20$ and $\alpha_2=0.40$	\label{f1}}
\end{figure}

\begin{figure}[H]
	\centering
	\includegraphics[height=2in, width=2.5in]{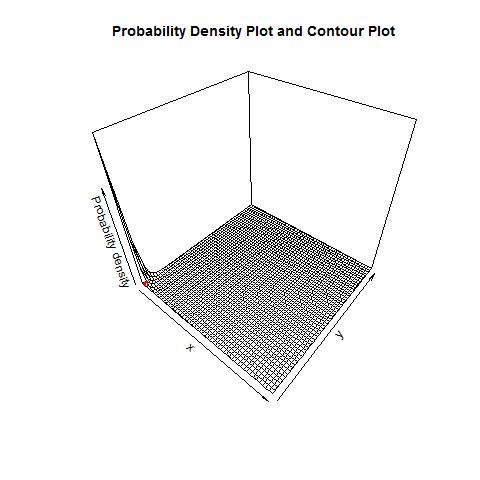}
	\caption{Probability density plot and Contour Plot for the parameter values $\theta = 0.80$, $\alpha_0=4$, $\alpha_1=5$ and $\alpha_2=10$ \label{f2}}
\end{figure}

  In this paper we are interested in parameter estimation of absolute continuous part only.  We consider the case when $\mu_1=0$, $\mu_2=0$, $\sigma_1=1$ and $\sigma_2=1$. We call the distribution as four parameter Geometric Block-Basu bivariate Pareto distribution and denote this as $G-BBBVPA(\theta, \alpha_0, \alpha_1, \alpha_2)$.  Then the joint density function of $(Y_1, Y_2)$ becomes,

\begin{align}
\label{2e1.}
g(y_1, y_2)&=
\begin{cases}
\frac{p\theta\alpha_1(\alpha_0 + \alpha_2)(1 + y_1)^{-\alpha_1 - 1}(1 + y_2)^{-\alpha_0-\alpha_2-1}\{1 + (1 - \theta)(1 + y_1)^{-\alpha_1}(1+ y_2)^{-\alpha_0-\alpha_2}\}}{\{1 - (1 - \theta)(1 + y_1)^{-\alpha_1}(1 + y_2)^{-\alpha_0-\alpha_2}\}^3}, \quad \text{if $y_1$ \textless $y_2$}
\\
\frac{p\theta(\alpha_0 + \alpha_1)\alpha_2(1 + y_1)^{-\alpha_0-\alpha_1-1}(1 + y_2)^{-\alpha_2-1}\{1 + (1 - \theta)(1 + y_1)^{-\alpha_0-\alpha_1}(1 + y_2)^{-\alpha_2}\}}{\{1 - (1 - \theta)(1 + y_1)^{-\alpha_0-\alpha_1}(1 + y_2)^{-\alpha_2}\}^3} ,\quad \text{if $y_1$ \textgreater $y_2$} 
\end{cases} 
\end{align}
where $p = \frac{\alpha_0 + \alpha_1 + \alpha_2}{\alpha_1 + \alpha_2}$.  The probability density plot and corresponding contour plot of different parameter sets are provided in Figure-\ref{f1} and Figure-\ref{f2} respectively.

\subsection{Marginal distributions}  The marginal distributions are easy to obtain from the above bivariate distribution which can be given by,

\begin{align}
\label{ma1}
\bar{G}(y_1) &=   p[\frac{\theta(1 + \frac{y_1 - \mu_1}{\sigma_1})^{-\alpha_0 - \alpha_1}}{1 - (1 - \theta)(1 + \frac{y_1 - \mu_1}{\sigma_1})^{-\alpha_0 - \alpha_1}} - (1 - \frac{1}{p})\frac{\theta(1 + \frac{y_1 - \mu_1}{\sigma_1})^{-(\alpha_0 + \alpha_1 + \alpha_2)}}{1 - (1 - \theta)(1 + \frac{y_1 - \mu_1}{\sigma_1})^{-(\alpha_0 + \alpha_1 + \alpha_2)}}],\nonumber\\ \quad \text{if $y_1 > \mu_1$}
\end{align}

\begin{align}
\label{ma2}
\bar{G}(y_2) &=   p[\frac{\theta(1 + \frac{y_2 - \mu_2}{\sigma_2})^{-\alpha_0 - \alpha_2}}{1 - (1 - \theta)(1 + \frac{y_2 - \mu_2}{\sigma_2})^{-\alpha_0 - \alpha_2}} - (1 - \frac{1}{p})\frac{\theta(1 + \frac{y_2 - \mu_2}{\sigma_2})^{-(\alpha_0 + \alpha_1 + \alpha_2)}}{1 - (1 - \theta)(1 + \frac{y_2 - \mu_2}{\sigma_2})^{-(\alpha_0 + \alpha_1 + \alpha_2)}}]],\nonumber\\ \quad \text{if $y_2 > \mu_2$}
\end{align}
where $p = \frac{\alpha_0 + \alpha_1 + \alpha_2}{\alpha_1 + \alpha_2}$.

\section{Parameter estimation through EM algorithm}

 Let us consider $I = \{(y_{11}, y_{21}), (y_{12}, y_{22}), \cdots, (y_{1m}, y_{2m})\}$, sample of size $m$ from four parameter G-BBBVPA distribution.  Let us also assume that $\mu_1$, $\mu_2$, $\sigma_1$ and $\sigma_2$ are known. 
 Now we use the following notation:

\begin{center}
 $ I_1 = \{i : y_{1i} < y_{2i} \}$, $ I_2 = \{ i : y_{1i} > y_{2i} \} $
\end{center}

  Also $m_1 = |I_1|$, $m_2 = |I_2|$, $m = m_1 + m_2$. 


 We assume this data corresponds to a fictitious singular distribution where cardinality of singular observation is $m_0$.  We form the usual EM in singular set up first.  Suppose we observe not only $(Y_1, Y_2)$ but also the corresponding $N$ value.  Hence the complete data would be of the form,
\begin{center}
$\{(y_{11}, y_{21}, n_1), (y_{12}, y_{22}, n_2) \cdots, (y_{1m^{*}}, y_{2m^{*}}, n_{m^{*}})\}$
\end{center}
Here $m^{*} = m_0 + m_1 + m_2$.

 We can imagine the following three independent hidden random variables corresponding to $(Y_1, Y_2)$,
\begin{center}
$\{ U_0 | N=n \} \sim PA(II)(0, 1, n\alpha_0)$
\end{center}
\begin{center}
$\{ U_1 | N=n \} \sim PA(II)(0, 1, n\alpha_1)$
\end{center}
\begin{center}
$\{ U_2 | N=n \} \sim PA(II)(0, 1, n\alpha_2)$
\end{center}
  Also it is well known that
\begin{center}
$\{ Y_1 | N=n \} = \min\{{\{ U_0 | N=n \}, \{ U_1 | N=n \}}\}$
\end{center}
\begin{center}
$\{ Y_2 | N=n \} = \min\{{\{ U_0 | N=n \}, \{ U_2 | N=n \}}\}$
\end{center}

Pseudo-likelihood function described in (\cite{DeyPaulGupta:2018}) can be obtained as :
\begin{align}
Q &= L_{pseudo}(\theta, \alpha_0, \alpha_1, \alpha_2) \nonumber\\& = \Big(\sum_{i\in I_0}\ln(a_i) + 2\sum_{i\in I_1}\ln(a_i) + 2\sum_{i\in I_2}\ln(a_i)\Big) + (m_0 + u_1 m_1 + w_1 m_2)\ln(\alpha_0) \nonumber\\& + (m_1 + w_2 m_2)\ln(\alpha_1) + (u_2 m_1 + m_2)\ln(\alpha_2)- \Big(\sum_{i\in I_0}\ln(1 + y_{1i}) + \sum_{i\in I_1\cup I_2}\ln(1 + y_{1i}) \nonumber\\& + \sum_{i\in I_1\cup I_2}\ln(1 + y_{2i})\Big) - \alpha_0\Big(\sum_{i\in I_0}a_i\ln(1 + y_{1i}) + \sum_{i\in I_1}a_i\ln(1 + y_{2i}) + \sum_{i\in I_2}a_i\ln(1 + y_{1i})\Big) \nonumber\\& - \alpha_1\Big(\sum_{i\in I_0}a_i\ln(1 + y_{1i})+ \sum_{i\in I_1}a_i\ln(1 + y_{1i})+\sum_{i\in I_2}a_i\ln(1 + y_{1i})\Big) - \alpha_2\Big(\sum_{i\in I_0}a_i\ln(1 + y_{1i})\nonumber\\& + \sum_{i\in I_1}a_i\ln(1 + y_{2i})+\sum_{i\in I_2}a_i\ln(1 + y_{2i})\Big) + m^{*}\ln(\frac{\theta}{1 - \theta}) + \ln(1 - \theta)\Big(\sum_{i\in I_0}a_i + \sum_{i\in I_1}a_i \nonumber\\& + \sum_{i\in I_2}a_i\Big)  \label{3e3}
\end{align}

 EM updates for the parameters $\theta$, $\alpha_0$, $\alpha_1$ and $\alpha_2$ are given as follows,
\begin{equation} 
\hat{\alpha}^{(t + 1)}_{0} = \frac{m_{0} + u^{(t)}_{1} m_{1} + w^{(t)}_{1} m_{2}}{\sum_{i\in I_0}a^{(t)}_i\ln(1+y_{1i})+\sum_{i\in I_1}a^{(t)}_i\ln(1+y_{2i})+\sum_{i\in I_2}a^{(t)}_i\ln(1+y_{1i})}. \label{3.e1}
\end{equation}
\begin{equation} 
\hat{\alpha}^{(t + 1)}_{1} = \frac{m_{1} + w^{(t)}_{2}m_{2}}{\sum_{i\in I_0}a^{(t)}_i\ln(1+y_{1i})+\sum_{i\in I_1}a^{(t)}_i\ln(1+y_{1i})+\sum_{i\in I_2}a^{(t)}_i\ln(1+y_{1i})} \label{3.e2}
\end{equation}
\begin{equation} \hat{\alpha}^{(t + 1)}_{2} = \frac{m_{2} + u^{(t)}_{2} m_{1}}{\sum_{i\in I_0}a^{(t)}_i\ln(1+y_{1i})+\sum_{i\in I_1}a^{(t)}_i\ln(1+y_{2i})+\sum_{i\in I_2}a^{(t)}_i\ln(1+y_{2i})} \label{3.e3}
\end{equation}
and
\begin{equation} \hat{\theta}^{(t + 1)} = \frac{m^{*}}{\sum_{i\in I_0}a^{(t)}_i+ \sum_{i\in I_1}a^{(t)}_i+\sum_{i\in I_2}a^{(t)}_i} \label{3.e4}
\end{equation}

where $a^{(t)}_i = E[N | y_{1i}, y_{2i}]$ is the conditional mean of $N$ given $Y_1 = y_1$ and $Y_2 = y_2$ at the step $t$ and $u^{(t)}_{1}, u^{(t)}_{2}, w^{(t)}_{1}$ and $w^{(t)}_{2}$ are posterior probabilities at time step $t$.

\noindent{\textbf{Important Issues and Suggested Solutions :}}

  We do not observe $m_0$ and each of the observation $U_0$ falling in $I_0$.  We also do not know $a_i$ when observations are in $I_0$. One of the straight solution is to replace all unknown quantities by its estimate $\tilde{m}_0$, $\tilde{b}_0$, $\tilde{a}_{0i}$.  So we replace $m_0$ and each of the observation $U_0$ falling in $I_0$ by it's estimates $\tilde{m}_0 = (m_1 + m_2)\frac{\alpha_0}{\alpha_1 + \alpha_2}$ and $\tilde{b}_0 = E[U_0 | U_0 < \min\{U_1, U_2\}] = \frac{1}{n(\alpha_0 + \alpha_1 + \alpha_2) - 1}$ , where $n$ should be replaced by $\tilde{a}_{0i}$.  We also use $\tilde{a}_{0i}$ as the estimate of $a_i$ falling in $I_0$.  \textbf{This method is valid when $\tilde{a}_{0i}(\alpha_0 + \alpha_1 + \alpha_2) > 1$}.  To make this a valid method for any range of parameters, we estimate $\log(1 + U_0)$ by $\tilde{b}^{*}_0 = E[\log(1 + U_0) | \log(1 + U_0) < \min\{\log(1 + U_1), \log(1 + U_2) \}] = \frac{1}{n(\alpha_0 + \alpha_1 + \alpha_2)}$ instead of estimating $U_0$.  Hence the modified EM estimates for the parameters $\theta$, $\alpha_0$, $\alpha_1$ and $\alpha_2$ of four parameter G-BBBVPA will look like,
\begin{equation}
\tilde{a}^{(t+1)}_{0i} = \frac{1 + (1 - \hat{\theta}^{(t)})e^{-1/\tilde{a}^{(t)}_{0i}}}{1 - (1 - \hat{\theta}^{(t)})e^{-1/\tilde{a}^{(t)}_{0i}}}\label{3..e1}
\end{equation}
\begin{equation}
\tilde{b}^{*(t+1)}_{0}=\frac{1}{\tilde{a}^{(t+1)}_{0i}\Big(\hat{\alpha}^{(t)}_{0} + \hat{\alpha}^{(t)}_{1} + \hat{\alpha}^{(t)}_{2}\Big)}\label{3..e2}
\end{equation}
\begin{equation} 
\hat{\alpha}^{(t + 1)}_{0} = \frac{\tilde{m}_0 + u^{(t)}_{1} m_{1} + w^{(t)}_{1} m_{2}}{\tilde{m}_0 \tilde{a}^{(t+1)}_{0i}\tilde{b}^{*(t+1)}_{0} + \sum_{i\in I_1} a^{(t)}_i\ln(1+y_{2i}) + \sum_{i\in I_2} a^{(t)}_i\ln(1+y_{1i})}. \label{3.e5}
\end{equation}
\begin{equation} 
\hat{\alpha}^{(t + 1)}_{1} = \frac{m_{1} + w^{(t)}_{2}m_{2}}{\tilde{m}_0 \tilde{a}^{(t+1)}_{0i}\tilde{b}^{*(t+1)}_{0} + \sum_{i\in I_1}a^{(t)}_i\ln(1 + y_{1i}) + \sum_{i\in I_2}a^{(t)}_i\ln(1 + y_{1i})} \label{3.e6}
\end{equation}
\begin{equation} \hat{\alpha}^{(t + 1)}_{2} = \frac{m_{2} + u^{(t)}_{2} m_{1}}{\tilde{m}_0 \tilde{a}^{(t+1)}_{0i}\tilde{b}^{*(t+1)}_{0} + \sum_{i\in I_1}a^{(t)}_i\ln(1 + y_{2i}) + \sum_{i\in I_2}a^{(t)}_i\ln(1 + y_{2i})} \label{3.e7}
\end{equation}
and
\begin{equation} \hat{\theta}^{(t + 1)} = \frac{\tilde{m}_0 + m_1 + m_2}{\tilde{m}_0 \tilde{a}^{(t+1)}_{0i}\tilde{b}^{*(t+1)}_{0} + \sum_{i\in I_1}a^{(t)}_i + \sum_{i\in I_2}a^{(t)}_i} \label{3.e8}
\end{equation}

Therefore the algorithmic steps of final working version can be given as :
   
\begin{algorithm}
\caption{Final modified EM procedure for four parameter Geometric absolute continuous Marshall Olkin bivariate Pareto distribution}
\begin{algorithmic}[1]
\STATE Set initial $\alpha_{0}$, $\alpha_{1}$, $\alpha_{2}$ and $\theta_{(i + 1)}$ and $a_{i}$.
 \WHILE{$\Delta Q/Q < tol$}
     \STATE Compute $\tilde{a}^{(t+1)}_{0i}$ using $\alpha^{(i)}_{0}$, $\alpha^{(i)}_{1}$, $\alpha^{(i)}_{2}$
     \STATE Compute $\tilde{b}^{*(t+1)}_{0i}$ using $\tilde{a}^{(t+1)}_{0i}$, $\alpha^{(i)}_{0}$, $\alpha^{(i)}_{1}$ and $\alpha^{(i)}_{2}$.
     \STATE Compute $u^{(i)}_{1}$, $u^{(i)}_{2}$, $w^{(i)}_{1}$, $w^{(i)}_{2}$, $\tilde{m}_0$ from $\alpha^{(i)}_{0}$, $\alpha^{(i)}_{1}$, $\alpha^{(i)}_{2}$ and $\tilde{a}^{(t+1)}_{0i}$.
     \STATE Update $\alpha^{(i+1)}_{0}$, $\alpha^{(i+1)}_{1}$, $\alpha^{(i+1)}_{2}$ and $\theta^{(i + 1)}$ using Equation (\ref{3.e5}), (\ref{3.e6}), (\ref{3.e7}) and (\ref{3.e8}).
     \STATE Calculate $Q$ for the new iterate.
\ENDWHILE
\end{algorithmic}
\label{algo3}
\end{algorithm}

\section{Bayesian Estimation}

    We use slice cum Gibbs sampler technique to calculate the bayes estimate.  Usual step out is easy to implement in case of informative prior which makes bayesian procedure also appealing for the practitioners.  At each Gibbs sampling step we plan to use slice sampling to generate the sample from the posterior.    

\subsection{Prior Assumption}

  We assume that $\alpha_0$, $\alpha_1$, and $\alpha_2$ are distributed according to the gamma distribution with shape parameters $k_i$ and scale parameters $\theta_i$, i.e.,
 \begin{eqnarray}
 \alpha_0 \sim \Gamma(k_0, \theta_0) \equiv \mathrm{Gamma}(k_0, \theta_0) \nonumber \\
 \alpha_1 \sim \Gamma(k_1, \theta_1) \equiv \mathrm{Gamma}(k_1, \theta_1) \nonumber \\
 \alpha_2 \sim \Gamma(k_2, \theta_2) \equiv \mathrm{Gamma}(k_2, \theta_2)
 \end{eqnarray}
 The probability density function of the Gamma Distribution is given by,
 \begin{align*}
 f_{\Gamma}(x; k, \theta) = \frac{1}{\Gamma(k)\theta^{k}}x^{k - 1}e^{-\frac{x}{\theta}}
 \end{align*} 
 Here $\Gamma(k)$ is the gamma function evaluated at $k$.

Further, we assume that the geometric parameter $\theta$ is distributed according to Beta distribution first kind with the parameters $a$ and $b$
\begin{align}
\pi(\theta) = \frac{1}{\beta(a, b)}\theta^{a-1}(1 - \theta)^{b-1}, \quad \theta \in [0, 1]
\end{align}

\subsection{Posterior Distribution}

  We known Bayes estimate of an unknown parameter under the squared error loss function is the posterior mean of the corresponding parameter.  But in this case it is not easy to calculate Bayes estimate of unknown model parameters  $\theta$, $\alpha_{0}$, $\alpha_{1}$ and $\alpha_{2}$ in closed form.  We propose slice cum Gibbs sampling procedure to generates sample from conditional posterior distribution.  The log full conditional posterior distributions of $\theta$, $\alpha_0$, $\alpha_1$ and $\alpha_2$ are given by,

\begin{align*}
\ln(\pi(\theta\mid \alpha_0, \alpha_1, \alpha_2, I)) &= L(\theta, \alpha_{0}, \alpha_{1}, \alpha_{2}) - \ln(\beta(a, b)) + (a-1)\ln\theta\\& + (b-1)\ln(1-\theta)
\end{align*}

\begin{align*}
\ln(\pi(\alpha_0\mid \theta, \alpha_1, \alpha_2, I)) &= L(\theta, \alpha_{0}, \alpha_{1}, \alpha_{2}) - \ln(\Gamma(k_0)) - k_0\ln(\theta_0)\\& +  (k_0-1)\ln\alpha_0-\frac{\alpha_0}{\theta_0}
\end{align*}

\begin{align*}
\ln(\pi(\alpha_1\mid \theta, \alpha_0, \alpha_2, I)) &= L(\theta, \alpha_{0}, \alpha_{1}, \alpha_{2}) - \ln(\Gamma(k_1)) - k_1\ln(\theta_1)\\& +  (k_1-1)\ln\alpha_1-\frac{\alpha_1}{\theta_1}
\end{align*}
and
\begin{align*}
\ln(\pi(\alpha_2\mid \theta, \alpha_0,\alpha_1, I)) &= L(\theta, \alpha_{0}, \alpha_{1}, \alpha_{2}) - \ln(\Gamma(k_2)) - k_2\ln(\theta_2)\\& +  (k_2-1)\ln\alpha_2-\frac{\alpha_2}{\theta_2}
\end{align*}
respectively.

\section{Numerical Analysis}

  We use the software R 3.5.0 to calculate all estimate.  The codes are run at IIT Guwahati computers with model : Intel(R) Core(TM) i5-6200U CPU 2.30GHz. The codes will be available on request to authors. Here we take two different set of parameters with two different sample size as $n=450$ and $n=1000$ in both of the estimation technique.

\subsection{EM Estimates:} For EM estimates, we generate sample from this distribution with different sample sizes ($n=300, 450, 1000$) for two different parameter sets to calculate average estimates (AE), mean squared error (MSE) and 95\% parametric bootstrap confidence interval (CI) of the parameters based on 1000 replications.  We also observe the average iteration number for each of the cases. We use different initial choice of the parameters for different parameter sets but it remains same for different sample size in EM algorithm. It is observed that all results are near about same with other initial choice within it's usual range.  We have used the stopping criterion as the relative difference of log-likelihood and pseudo log-likelihood at each of the iteration with stopping tolerance limit as $0.00001$. All MLE results are shown in Table-\ref{5t1}, Table-\ref{5t2}, Table-\ref{5t3}, Table-\ref{5t4}, Table-\ref{5t5} and Table-\ref{5t6} respectively.  The simulation provides average estimates closer to the original parameter with low MSEs which indicates that the method works quite well even for moderate sample size.
\begin{table}[H]
	\centering
	\begin{adjustbox}{width=1.0\textwidth}
		{\begin{tabular}[l]{@{}lcccc}\hline
				EM Algorithm & & & & \\ \hline
				Parameter Sets & $\theta$ & $\alpha_0$ & $\alpha_1$ & $\alpha_2$ \\ 
				Average Estimates & 0.1880  & 0.0896 & 0.1923 & 0.3802 \\
                Mean Squared Error & 0.0031 & 0.0022 & 0.0041 & 0.0131\\
                Confidence Interval & [0.0824 0.2825] & [0.0018 0.1976] & [0.0863 0.3134] & [0.1771 0.5822] \\ \hline 
                Average Number of Iteration & &  & 251 &\\ \hline
		\end{tabular}}
	\end{adjustbox}
	\caption{Average Estimates (AE), Mean Square Errors (MSE) and Confidence Intervals (CI) of four parameter G-BBBPA with parameters $\theta=0.20$, $\alpha_0=0.1$, $\alpha_1=0.2$, $\alpha_2=0.4$ and sample size $n=300$}
	\label{5t1}
\end{table}

\begin{table}[H]
\centering
\begin{adjustbox}{width=1.0\textwidth}
{\begin{tabular}[l]{@{}lcccc}\hline
EM Algorithm & & & & \\ \hline
 Parameter Sets & $\theta$ & $\alpha_0$ & $\alpha_1$ & $\alpha_2$ \\ 
 Average Estimates & 0.1891  & 0.0910 & 0.1926 & 0.3817 \\
 Mean Squared Error & 0.0025 & 0.0017 & 0.0030 & 0.0100\\
 Confidence Interval & [0.0809 0.2647] & [0.0230 0.1783] & [0.0868 0.2861] & [0.1759 0.5431] \\ \hline 
 Average Number of Iteration & &  & 238 &\\ \hline
\end{tabular}}
\end{adjustbox}
\caption{Average Estimates (AE), Mean Square Errors (MSE) and Confidence Intervals (CI) of four parameter G-BBBPA with parameters $\theta=0.20$, $\alpha_0=0.1$, $\alpha_1=0.2$, $\alpha_2=0.4$ and sample size $n=450$}
\label{5t2}
\end{table}

\begin{table}[H]
\centering
\begin{adjustbox}{width=1.0\textwidth}
{\begin{tabular}[l]{@{}lcccc}\hline
EM Algorithm & & & & \\ \hline
 Parameter Sets & $\theta$ & $\alpha_0$ & $\alpha_1$ & $\alpha_2$ \\ 
 Average Estimates &  0.1926  & 0.0939 & 0.1948 & 0.3884 \\
 Mean Squared Error & 0.0014 & 0.0008 & 0.0016 & 0.0056\\
 Confidence Interval & [0.0812 0.2464] & [0.0428 0.1480] & [0.0870 0.2581] & [0.1734 0.4929] \\ \hline 
Average Number of Iteration & &  & 225 &\\ \hline
\end{tabular}}
\end{adjustbox}
\caption{Average Estimates (AE), Mean Square Errors (MSE) and Confidence Intervals (CI) of four parameter G-BBBVPA with parameters $\theta=0.20$, $\alpha_0=0.1$, $\alpha_1=0.2$, $\alpha_2=0.4$ and sample size $n=1000$}
\label{5t3}
\end{table}

\begin{table}[H]
	\centering
	\begin{adjustbox}{width=1.0\textwidth}
		{\begin{tabular}[l]{@{}lcccc}\hline
				EM Algorithm & & & & \\ \hline
				Parameter Sets & $\theta$ & $\alpha_0$ & $\alpha_1$ & $\alpha_2$ \\ 
				Average Estimates & 0.8091  & 4.0329 & 5.0769 & 10.0932 \\
                Mean Squared Error & 0.0085 & 1.8467 & 0.9753 & 2.1252\\
                Confidence Interval & [0.6355 0.9992] & [1.4026 6.7785] & [3.3735 7.1640] & [7.3086 13.0668] \\ \hline 
                Average Number of Iteration & &  & 316 &\\ \hline
		\end{tabular}}
	\end{adjustbox}
	\caption{Average Estimates (AE), Mean Square Errors (MSE) and Confidence Intervals (CI) of four parameter G-BBBVPA with parameters $\theta=0.80$, $\alpha_0=4$, $\alpha_1=5$, $\alpha_2=10$ and sample size $n=300$}
	\label{5t4}
\end{table}

\begin{table}[H]
\centering
\begin{adjustbox}{width=1.0\textwidth}
{\begin{tabular}[l]{@{}lcccc}\hline
EM Algorithm & & & & \\ \hline
 Parameter Sets & $\theta$ & $\alpha_0$ & $\alpha_1$ & $\alpha_2$ \\ 
 Average Estimates & 0.8062  & 4.0274 & 5.0526 & 10.0416 \\
 Mean Squared Error & 0.0062 & 1.1381 & 0.6251 & 1.3914\\
 Confidence Interval & [0.6628 0.9678] & [1.8657 5.9532] & [3.6413 6.6936] & [7.7724 12.4351] \\ \hline 
 Average Number of Iteration & &  & 285 &\\ \hline
\end{tabular}}
\end{adjustbox}
\caption{Average Estimates (AE), Mean Square Errors (MSE) and Confidence Intervals (CI) of four parameter G-BBBVPA with parameters $\theta=0.80$, $\alpha_0=4$, $\alpha_1=5$, $\alpha_2=10$ and sample size $n=450$}
\label{5t5}
\end{table}

\begin{table}[H]
\centering
\begin{adjustbox}{width=1.0\textwidth}
{\begin{tabular}[l]{@{}lcccc}\hline
EM Algorithm & & & & \\ \hline
 Parameter Sets & $\theta$ & $\alpha_0$ & $\alpha_1$ & $\alpha_2$ \\ 
 Average Estimates & 0.8016  & 4.0276 & 5.0191 &  5.0191 \\
 Mean Squared Error & 0.0029 & 0.5228 &  0.2727 & 0.6100\\
 Confidence Interval & [0.70491 0.9051] & [2.5286 5.4450] & [4.1197 6.0914] & [8.5197 11.5006] \\ \hline 
 Average Number of Iteration & &  & 253 &\\ \hline
\end{tabular}}
\end{adjustbox}
\caption{Average Estimates (AE), Mean Square Errors (MSE) and Confidence Intervals (CI) of four parameter G-BBBVPA with parameters $\theta=0.80$, $\alpha_0=4$, $\alpha_1=5$, $\alpha_2=10$ and sample size $n=1000$}
\label{5t6}
\end{table}

\subsection*{For Bayesian Estimation:}

   We compute the average bayesian estimates (ABE) of unknown parameters and also the associated mean squared error (MSE), credible intervals (CI) and coverage probability (CP) of CIs based on proper priors with fixed hyper-parameters. Although the code is done based on one particular set of hyper-parameters, it may work for any set of hyper-parameters.  In step out method of slice sampling we choose our width as one.  However we cross check that the algorithm works for both larger and smaller choices of width.  The confidence intervals are constructed directly using R package coda.  All bayesian results are shown in Table-\ref{5t7}, Table-\ref{5t8}, Table-\ref{5t9},  Table-\ref{5t10}, Table-\ref{5t11} and Table-\ref{5t12}  respectively.  Results indicate that the method used in this case works really well even for moderate sample size.  We use the following hyper-parameters $a=0.70$, $b=0.75$, $k_0=0.70$, $\theta_0=0.75$, $k_1=0.70$, $\theta_1=0.75$, $k_2=0.70$ and $\theta_2=0.75$ for Gamma priors.  We observe in simulation experiment that the method works for multiple different choices of hyper-parameters indicating that the algorithm is independent of the choice of the hyper-parameters.
\begin{table}[H]
	\centering
	\begin{adjustbox}{width=1.0\textwidth}
		{\begin{tabular}[l]{@{}lcccc}\hline
				Slice-cum-Gibbs & & & & \\ \hline
				Gamma Prior &  & & & \\ \hline
				Parameter Sets & $\theta$ & $\alpha_0$ & $\alpha_1$ & $\alpha_2$ \\ 
				Starting Value & 0.4794 & 0.8654 & 0.7781 & 0.5386 \\
				Average Estimates & 0.2142  & 0.1001 & 0.2178 & 0.4252 \\
				Mean Squared Error & 0.0014 & 0.0023 & 0.0020 & 0.0055\\
				Credible Interval & [0.1882 0.3577] & [0.00001 0.1889] & [0.1684 0.37901] & [0.3296 0.6382] \\
				Coverage Probabilities & 0.95 & 0.915 & 0.96 & 0.965 \\ \hline
		\end{tabular}}
	\end{adjustbox}
	\caption{Average Bayesian Estimates (ABE), Mean Square Errors (MSE), Credible Intervals (CI) and Coverage Probabilities (CP) of the parameters of four parameter G-BBBVPA with parameters $\theta=0.20$, $\alpha_0=0.10$, $\alpha_1=0.20$, $\alpha_2=0.40$ and sample size $n=300$}
	\label{5t7}
\end{table}

\begin{table}[H]
	\centering
	\begin{adjustbox}{width=1.0\textwidth}
		{\begin{tabular}[l]{@{}lcccc}\hline
				Slice-cum-Gibbs & & & & \\ \hline
				Gamma Prior &  & & & \\ \hline
				Parameter Sets & $\theta$ & $\alpha_0$ & $\alpha_1$ & $\alpha_2$ \\ 
				Starting Value & 0.4794 & 0.8654 & 0.7781 & 0.5386 \\
                Average Estimates & 0.2074  & 0.0987 & 0.2098 & 0.4130 \\
                Mean Squared Error & 0.0008 & 0.0012 & 0.0015 & 0.0042\\
                Credible Interval & [0.1897 0.3264] & [0.00001 0.1607] & [0.1908 0.3614] & [0.3281 0.5755] \\
                Coverage Probabilities & 0.96 & 0.975 & 0.955 & 0.93 \\ \hline
		\end{tabular}}
	\end{adjustbox}
	\caption{Average Bayesian Estimates (ABE), Mean Square Errors (MSE), Credible Intervals (CI) and Coverage Probabilities (CP) of the parameters of four parameter G-BBBVPA with parameters $\theta=0.20$, $\alpha_0=0.10$, $\alpha_1=0.20$, $\alpha_2=0.40$ and sample size $n=450$}
	\label{5t8}
\end{table}

\begin{table}[H]
	\centering
	\begin{adjustbox}{width=1.0\textwidth}
		{\begin{tabular}[l]{@{}lcccc}\hline
				Slice-cum-Gibbs & & & & \\ \hline
				Gamma Prior &  & & & \\ \hline
				Parameter Sets & $\theta$ & $\alpha_0$ & $\alpha_1$ & $\alpha_2$ \\ 
				Starting Value & 0.4794 & 0.8654 & 0.7781 & 0.5386 \\
                Average Estimates & 0.2045  & 0.0970 & 0.2067& 0.4086 \\
                Mean Squared Error & 0.0004 & 0.0008 & 0.0007 & 0.0020\\
                Credible Interval & [0.1933 0.2797] & [0.00829 0.1211] & [0.1880 0.3019] & [0.3691 0.5460] \\
                Coverage Probabilities & 0.935 & 0.93 & 0.945 & 0.935 \\ \hline
		\end{tabular}}
	\end{adjustbox}
	\caption{Average Bayesian Estimates (ABE), Mean Square Errors (MSE), Credible Intervals (CI) and Coverage Probabilities (CP) of the parameters of four parameter G-BBBVPA with parameters $\theta=0.20$, $\alpha_0=0.10$, $\alpha_1=0.20$, $\alpha_2=0.40$ and sample size $n=1000$}
	\label{5t9}
\end{table}

\begin{table}[H]
	\centering
	\begin{adjustbox}{width=1.0\textwidth}
		{\begin{tabular}[l]{@{}lcccc}\hline
				Slice-cum-Gibbs & & & & \\ \hline
				Gamma Prior &  & & & \\ \hline
				Parameter Sets & $\theta$ & $\alpha_0$ & $\alpha_1$ & $\alpha_2$ \\ 
				Starting Value & 0.4794 & 0.8654 & 0.7781 & 0.5386 \\
				Average Estimates & 0.6830  & 4.3083 & 3.9428 & 7.8710 \\
				Mean Squared Error & 0.0197 & 1.2093 & 1.6216 & 5.9848\\
				Credible Interval & [0.6231 0.9841] & [3.9782 9.0116] & [2.0142 4.8029] & [4.1524 9.1037] \\
				Coverage Probabilities & 0.75 & 0.96 & 0.68 & 0.585 \\ \hline
		\end{tabular}}
	\end{adjustbox}
	\caption{Average Bayesian Estimates (ABE), Mean Square Errors (MSE), Credible Intervals (CI) and Coverage Probabilities (CP) of the parameters of four parameter G-BBBVPA with parameters $\theta=0.80$, $\alpha_0=4$, $\alpha_1=5$, $\alpha_2=10$ and sample size $n=300$}
	\label{5t10}
\end{table}

\begin{table}[H]
	\centering
	\begin{adjustbox}{width=1.0\textwidth}
		{\begin{tabular}[l]{@{}lcccc}\hline
				Slice-cum-Gibbs & & & & \\ \hline
				Gamma Prior &  & & & \\ \hline
				Parameter Sets & $\theta$ & $\alpha_0$ & $\alpha_1$ & $\alpha_2$ \\ 
				Starting Value & 0.4794 & 0.8654 & 0.7781 & 0.5386 \\
                Average Estimates & 0.7198  & 4.1272 & 4.3771 & 8.6477 \\
                Mean Squared Error & 0.0120 & 0.9798 & 0.8628 & 2.9695\\
                Credible Interval & [0.7219 0.9996] & [3.2084 7.2971] & [3.0594 5.8116] & [6.6313 11.0700] \\
                Coverage Probabilities & 0.74 & 0.955 & 0.79 & 0.735 \\ \hline
		\end{tabular}}
	\end{adjustbox}
	\caption{Average Bayesian Estimates (ABE), Mean Square Errors (MSE), Credible Intervals (CI) and Coverage Probabilities (CP) of the parameters of four parameter G-BBBVPA with parameters $\theta=0.80$, $\alpha_0=4$, $\alpha_1=5$, $\alpha_2=10$ and sample size $n=450$}
	\label{5t11}
\end{table}

\begin{table}[H]
	\centering
	\begin{adjustbox}{width=1.0\textwidth}
		{\begin{tabular}[l]{@{}lcccc}\hline
				Slice-cum-Gibbs & & & & \\ \hline
				Gamma Prior &  & & & \\ \hline
				Parameter Sets & $\theta$ & $\alpha_0$ & $\alpha_1$ & $\alpha_2$ \\ 
				Starting Value & 0.4794 & 0.8654 & 0.7781 & 0.5386 \\
                Average Estimates & 0.7623  & 4.0852 & 4.7265 & 9.3589 \\
                Mean Squared Error & 0.0043 & 0.4909 & 0.3335 & 0.9998\\
                Credible Interval & [0.6966 0.9068] & [3.7685  6.3895] & [3.1651 4.7518] & [7.6523 10.3458] \\
                Coverage Probabilities & 0.86 & 0.96 & 0.895 & 0.82 \\ \hline
		\end{tabular}}
	\end{adjustbox}
	\caption{Average Bayesian Estimates (ABE), Mean Square Errors (MSE), Credible Intervals (CI) and Coverage Probabilities (CP) of the parameters of four parameter G-BBBVPA with parameters $\theta=0.80$, $\alpha_0=4$, $\alpha_1=5$, $\alpha_2=10$ and sample size $n=1000$}
	\label{5t12}
\end{table}

\section{Data Analysis}

\begin{figure}[H]
	\centering
		\includegraphics[height=2in, width=2.5in]{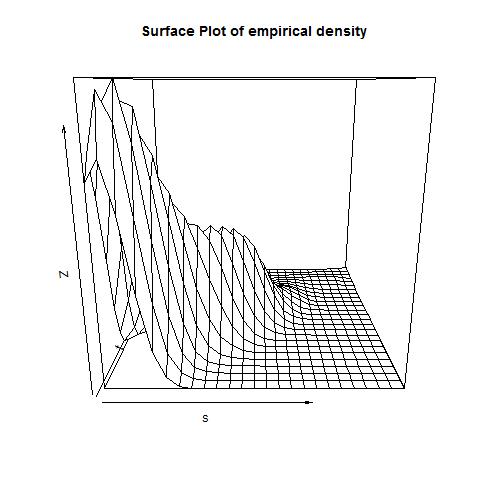}
	\caption{Two dimensional density plots of the transformed dataset \label{df11}}
\end{figure}

\begin{figure}[H]
	\centering
	\subfigure[For $X_{1}$]{
		\includegraphics[height=2in]{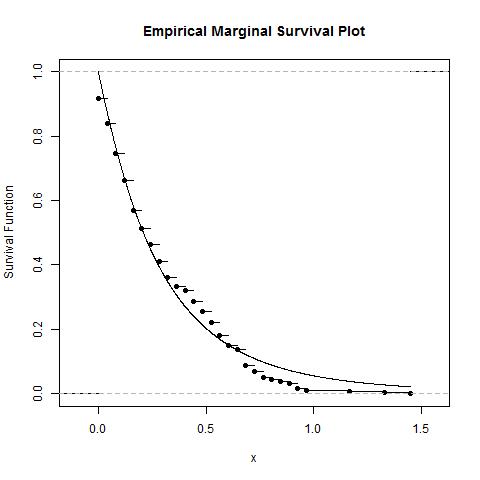}}
	\qquad
	\subfigure[For $X_{2}$]{
		\includegraphics[height=2in]{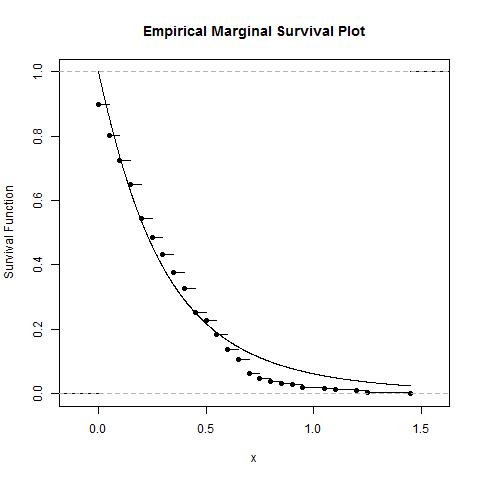}}
	\caption{Survival plots for two marginals through EM algorithm of the transformed dataset \label{df1}}
\end{figure}

\begin{figure}[H]
	\centering
	\subfigure[For $X_{1}$]{
		\includegraphics[height=2in]{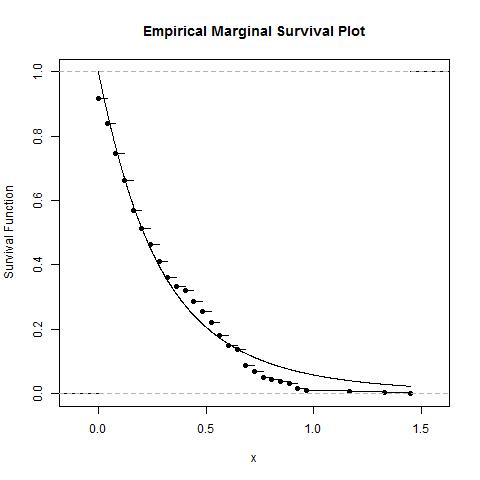}}
	\qquad
	\subfigure[For $X_{2}$]{
		\includegraphics[height=2in]{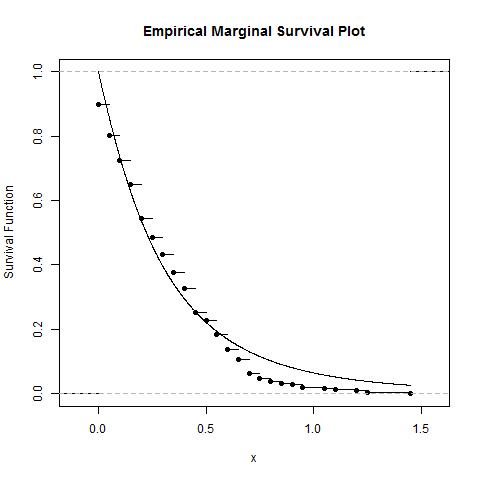}}
	\caption{Survival plots for two marginals through Bayesian Estimation of the transformed dataset \label{df2}}
\end{figure}

  We use the same data set which is used by Paul Dey and Kundu (2018) for data analysis of three parameter BB-BVPA distribution.  The data set is taken from \url{https://archive.ics.uci.edu/ml/machine-learning-databases}.  The age of abalone is determined by cutting the shell through the cone, staining it, and counting the number of rings through a microscope. The data set contains related measurements.  We extract a part of the data for bivariate modeling.  We consider only measurements related to female population where one of the variable is Length as Longest shell measurement and other variable is Diameter which is perpendicular to length. We use peak over threshold method on this data set.  It is observed that the transform data set doesn't have any singular component. So instead of modeling with BBBVPA, we can plan to choose more generalized/flexible Geometric BBBVPA as one of the possible distributional assumption.  We fit the empirical survival functions with the marginals of this distribution whose parameters are obtained from the both of proposed EM algorithm and Bayesian estimation. For the both marginals it has a good fit which are shown in Figure-\ref{df1} and Figure-\ref{df2} respectively for EM algorithms.  We also verify our assumption by plotting empirical two dimensional density plot of this data which is shown in Figure-\ref{df11}.

\paragraph*{For EM algorithm:}

 The EM estimates of the parameters of four parameter BBBVPA based on transform data are $\theta = 0.9999$, $\alpha_0=3.1388$, $\alpha_1=1.7324$ and $\alpha_2=1.5920$.   Average estimates, mean square errors, confidence intervals and average number of iterations based on parametric bootstrap  are available in Table-\ref{6t1}. 

\begin{table}[H]
\centering
\begin{adjustbox}{width=1.0\textwidth}
{\begin{tabular}[l]{@{}lcccc}\hline
EM Algorithm & & & & \\ \hline
 Parameter Sets & $\theta$ & $\alpha_0$ & $\alpha_1$ & $\alpha_2$ \\ 
 Average Estimates & 0.9649  & 3.1298 & 1.6828 & 1.5462 \\
 Mean Squared Error & 0.0040 & 0.3873 & 0.2126 & 0.1877\\
 Confidence Interval & [0.8226 0.9999] & [1.9728 4.3470] & [0.8174 2.6112] & [0.7293 2.3890] \\ \hline 
 Average Number of Iteration & &  & 1036 &\\ \hline
\end{tabular}}
\end{adjustbox}
\caption{Average Estimates (AE), Mean Square Errors (MSE) and Confidence Intervals (CI) of four parameter G-BBBVPA with parameters$\theta=0.9999$, $\alpha_0=3.1388$, $\alpha_1=1.7324$, $\alpha_2=1.5921$ and sample size $n=329$}
\label{6t1}
\end{table}

\paragraph*{For Bayesian Estimation:}

  The Bayesian estimates of the parameters based on this data are $\theta = 0.9852$, $\alpha_0=3.5395$, $\alpha_1=1.3782$ and $\alpha_2=1.2632$.  We also calculate average bayesian estimates, mean squared error, credible intervals and coverage probability of the credible intervals using simulation technique which is available in Table-\ref{6t2}.

\begin{table}[H]
	\centering
	\begin{adjustbox}{width=1.0\textwidth}
		{\begin{tabular}[l]{@{}lcccc}\hline
				Slice-cum-Gibbs & & & & \\ \hline
				Gamma Prior &  & & & \\ \hline
				Parameter Sets & $\theta$ & $\alpha_0$ & $\alpha_1$ & $\alpha_2$ \\ 
				Starting Value & 0.4794 & 0.8654 & 0.7781 & 0.5386 \\
                Average Estimates & 0.9067  & 3.3691 & 1.3061 &  1.1967 \\
                Mean Squared Error & 0.0082 & 0.3280 & 0.1383 & 0.1175 \\
                Credible Interval & [0.8637 1.000] & [3.1685 5.5302] & [0.0247 1.5164] & [0.0213 1.4281] \\
                Coverage Probabilities & 0.975 & 0.95 & 0.965 & 0.955 \\ \hline
		\end{tabular}}
	\end{adjustbox}
	\caption{Average Bayesian Estimates (ABE), Mean Square Errors (MSE), Credible Intervals (CI) and Coverage Probabilities (CP) of the parameters of four parameter G-BBBVPA with parameters $\theta=0.9852$, $\alpha_0=3.5395$, $\alpha_1=1.3781$, $\alpha_2=1.2632$ and sample size $n=329$}
	\label{6t2}
\end{table}

\section{Conclusion}  We observe that it is a more flexible model than three parameter absolute continuous Marshall-Olkin bivariate Pareto distribution.  We propose different modification while implementing the EM algorithm. Bayesian analysis through Slice cum Gibbs sampler is used for estimation of the parameters too.  Modeling the data through G-MOBVPA may have more appeal for the practitioner when there is no equality in components of the model.  However statistical inference for this model is much more difficult with location and scale due to its discontinuous nature of likelihood with respect to the parameters.  A different alternative version of this Geometric distribution can also be proposed in the same direction.  The work is on progress. 

\bibliographystyle{chicago}

\bibliography{bibliography-comm}

\end{document}